\documentclass[12pt,a4,twocolumn]{article}
\usepackage[utf8]{inputenc}
\usepackage[OT4]{fontenc}
\usepackage{amsmath}
\usepackage{amssymb}
\usepackage{graphicx}
\usepackage{multirow}

\begin{document}
\title{A circular polymer chain in a gel - the reduction of the state space}
\author{Malgorzata J. Krawczyk\\
Faculty of Physics and Applied Computer Science,\\
AGH University of Science and Technology,\\
al. Mickiewicza 30, 30-059 Krakow, Poland\\
e-mail: gos@fatcat.ftj.agh.edu.pl
}
\date{January 24, 2012}

%\keywords{state space, network of states, reduction of the system size, exact enumeration for weighted networks}
\maketitle
\begin{abstract}
The state space of a polymer molecule is analysed. We show how the size of the state space can be reduced on the basis of symmetry. In the reduced state space, the probability of a new state (termed below as class) is equal to the number of old states represented by the new state multiplied by the probability of each old state. As an application, the electrophoretic motion of the molecule in gel is considered. We discuss the influence of the gel medium and of external field on the molecule states, with absorbing states of hooked molecules playing a major role. We show that in the case of strong fields both the velocity and the diffusion coefficient decrease with field. Finally, we evaluate the time of relaxation to and from the absorbing states.  This is done with a continuous version of the exact enumeration method for weighted networks. 
\end{abstract}

\section{Introduction}

Computational tasks in biology, sociology, physics and many other areas meet in most cases limitations of computer memory and time, which limits the size of the modelled systems.  It is worthwhile to look for methods how to reduce the size of an analysed system without loss of information. Recently we have shown \cite{mk1,mk2} that when some symmetries are present, the size of the state space can be significantly reduced. In our previous papers we applied the new method to spin lattices \cite{mk1} and traffic systems \cite{mk2}. In both cases the same formalism was used. The idea is to group similar states on the basis of their symmetry. The procedure preserves information on the static probabilities of all states.\\

The same idea is now applied to the state space formed by a set of possible conformations of a circular polymer molecule. A standard model description of this kind of molecules is the repton model \cite{deg,new}. The main trick of this model is to divide the whole molecule into parts, so-called reptons. We are interested in all possible conformations of a molecule, assuming that reptons are placed at nodes of a regular, square lattice (another basic assumption of the repton model). This means that the distance between two neighbouring reptons can be equal to $0$ or $1$ in units of lattice constant. In other words, two neighbouring reptons can be placed in the same position or in an assumed, constant distance.  A symmetry of the space will also be visible in molecule conformations, which allows for a reduction of the size of the state space of the system.\\

Modelling of conformations of polymer molecules and transitions between them finds applications, e.g. in electrophoretic transport in gel. This process is commonly used in biology for separating molecules, such as DNA or peptides, according to their size. The process is based on the motion of charged polymer molecules in an external electric field in a porous medium. In the theory of electrophoresis one can distinguish different regimes of molecule migration through the media, which depend on the molecule length, the gel density and the electric field strength. In the case of weak fields the process of molecule displacement is a result of thermal diffusion and motion caused by the applied electric field. In the case of strong fields, the picture changes significantly as thermal shifts become more rare. It may lead molecules to hook on gel fibres, which reduces their speed or even immobilize them \cite{el1,el2}.\\

The paper is organised as follows: First, the state space of the molecule conformations is presented. We show how the symmetry of the circular molecule allows the reduction of the size of the system. Then we present classes of states for the original and the reduced system. In the following section we show how the state space is changed because of the presence of the porous medium and an external electric field. Section \ref{dyn} is devoted to the discussion of the probability of states and the probabilities of particular transitions between states in an external field. The following section presents the results of our calculations of the velocity of molecules, the diffusion coefficient and the relaxation time of the system as dependent on the applied field. In the case of the relaxation time, the reduction of the state space leads to approximate solutions. The final section is devoted to discussion.

\section{Analysed system}
\subsection{State space}
The state space is formed from the set of all possible conformations of a circular polymer molecule in the repton model. We also assume that the reptons are indistinguishable. This means in particular that a shift of the whole molecule along its contour does not lead to a different state. We note that the lattice cells are distinguishable. However, a translation of the whole molecule in the lattice does not generate a new state. With these assumptions, we find all possible conformations of a molecule of a given length. The length of the molecule is expressed by its number of reptons $N$.\\

\subsection{Network of states}
\begin{figure}[!hptb]
\begin{center}
\includegraphics[width=.3\textwidth, angle=270]{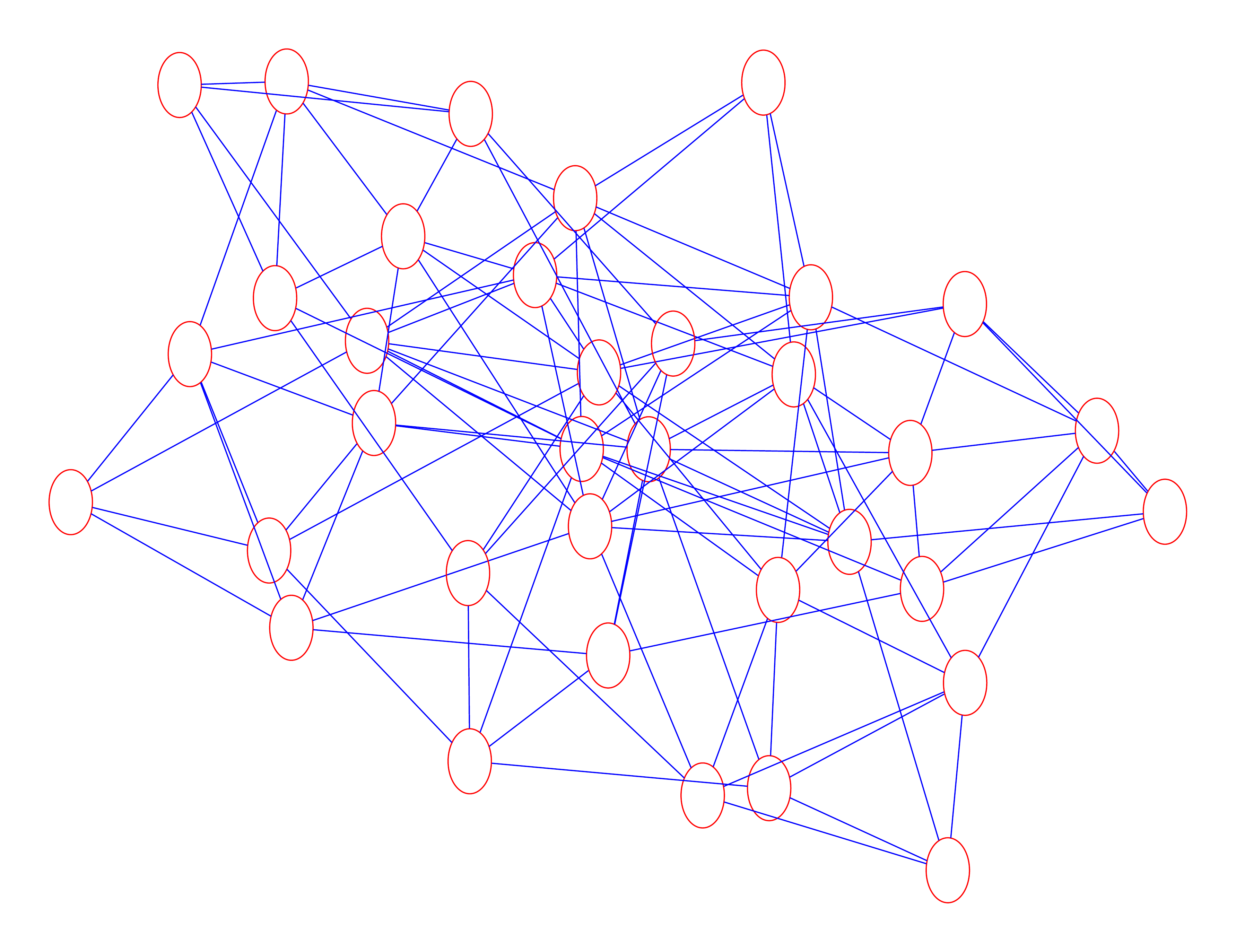}
\caption{Network of states for 5-repton long molecule.}
\label{net}
\end{center}
\end{figure}

A transition between two molecule conformations is possible if it is connected with the displacement of one repton from one lattice node to another one. One can think about states as about nodes of some network of states, with edges determined by possible one-repton moves between the states. In this terminology the number of connections of a given state to other states is equivalent to the node degree. The network obtained for a circular molecule of five reptons is presented in Fig.\ref{net}. 

\begin{figure}[!hptb]
\begin{center}
\includegraphics[width=.3\textwidth, angle=0]{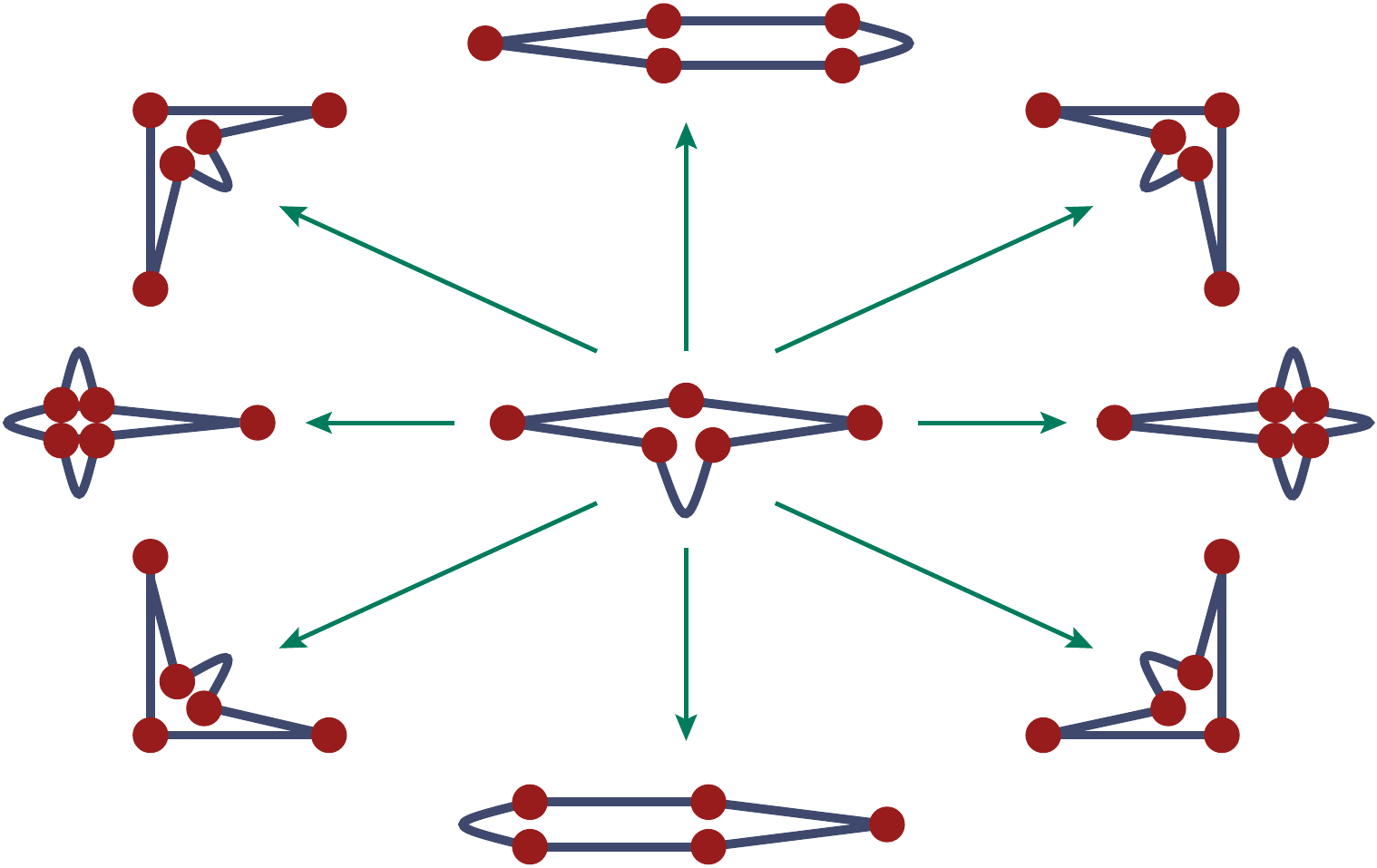}
\caption{Examples of possible one-repton transitions between states.}
\label{prze}
\end{center}
\end{figure}

In Fig.\ref{prze} all possible transitions from one example state are presented for a molecule of 5 reptons. All of them are connected with the change of the position of one repton, and the distance between two neighbouring reptons is equal to the lattice constant or to $0$.

\subsection{Classes}
As points in the space are distinguishable, molecule conformations which are mirror reflections are seen as different states. However, in fact the distinction between those states is illusory. Existing symmetry of the space causes that all states which can be obtained one from another as a result of reflection can be treated as one class. As a consequence, the effective size of the network of states can be reduced.

\begin{figure}[!hptb]
\begin{center}
\includegraphics[width=.3\textwidth, angle=0]{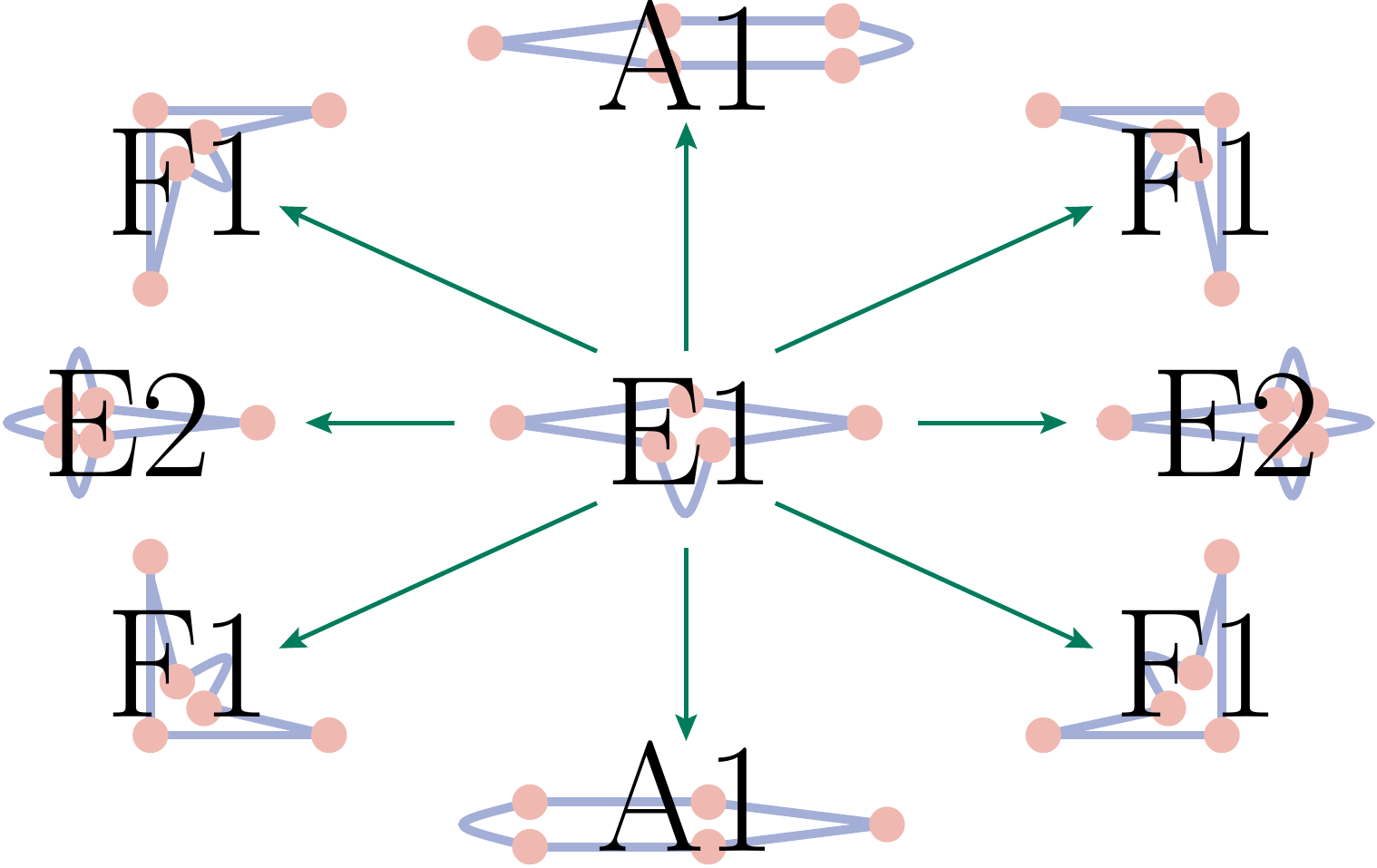}
\caption{Examples of possible one-repton transitions between classes.}
\label{przeS}
\end{center}
\end{figure}

Formally the procedure of reduction of the size of the state space of the system is performed as follows: The quantity which characterises each state is the number of transitions between this state and other states of a system. The number of possible transitions from a given state and the set of states it can be transformed to determine a class to which the state is classified. Each conformation of the molecule is then represented by the class it is classified to (see Fig.\ref{przeS}).

The inspection of all possible transitions leads to the division of the whole set of states into classes. The size of a state space and the number of classes depend on the length of the molecule. Obtained results are presented in Tab.\ref{classes}. As it can be seen in these data, the number of possible states grows significantly with the number of reptons, so in the following part of the paper we will work with a molecule of 5 reptons. 

\begin{table}
{\tiny{
\begin{tabular}{c|c|c}
\textsc{$N$}&\textsc{number of states}&\textsc{number of classes}\\\hline
$5$&$35$&$9$\\
$6$&$123$&$31$\\
$7$&$359$&$62$\\
$8$&$1363$&$241$\\
$9$&$4\;891$&$702$
\end{tabular}
}}
\caption{Number of states and number of classes for molecules of different lengths $N$ (expressed via the number of reptons)}
\label{classes}
\end{table}

A full classification of the states of a circular molecule of 5 reptons is presented in Tab.\ref{classesA}. The symbol in the first column indicates a class. In the second column classes of states are listed which are attainable from the states classified to the class specified in the first column. The third column covers information about the number of possible transitions, and in the last column the number of states which are classified to a given class is given.

\begin{table}
{\tiny{
\begin{tabular}{c|*{9}{l}|c|c}
\textsc{class}&\multicolumn{9}{c|}{\textsc{classes given class can be transformed to}}&\textsc{degree}&\textsc{\#states}\\\hline
A1&B1&B1&C1&E1&&&&&&4&4\\
A2&E2&E2&E2&E2&&&&&&4&1\\
B1&A1&B2&C1&D1&F1&&&&&5&8\\
B2&B1&B1&B2&B2&F1&&&&&5&4\\
C1&A1&B1&B1&C1&D1&E2&&&&6&4\\
D1&B1&B1&C1&D1&E2&F1&F1&&&7&4\\
E1&A1&A1&E2&E2&F1&F1&F1&F1&&8&2\\
E2&A2&C1&D1&E1&E2&E2&F1&F1&&8&4\\
F1&B1&B1&B2&D1&D1&E1&E1&E2&E2&9&4\\
\end{tabular}
}}
\caption{Classes identified for the 5-repton long molecule}
\label{classesA}
\end{table}

\begin{figure}[!hptb]
\begin{center}
\includegraphics[width=.3\textwidth, angle=0]{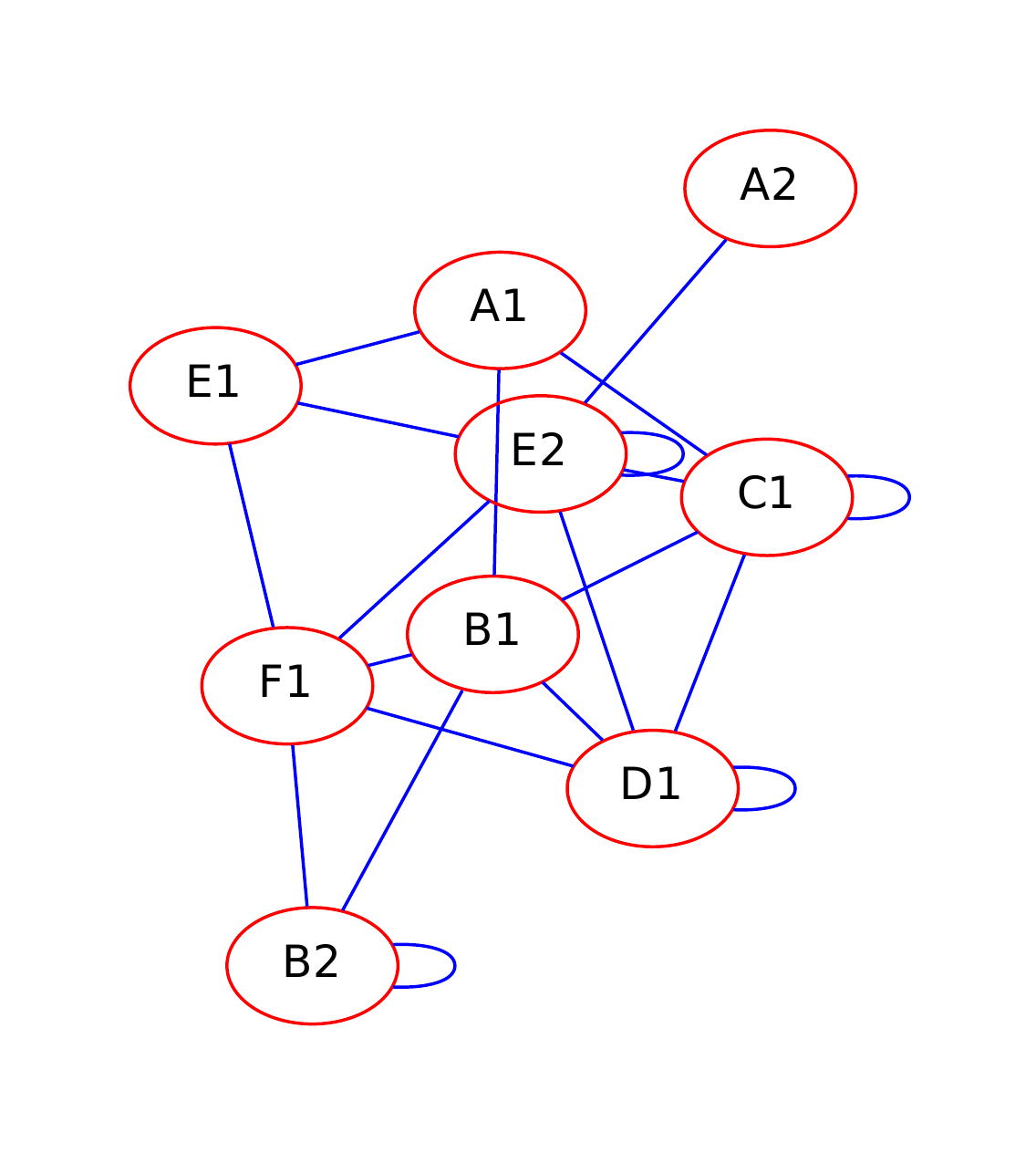}
\caption{Network of classes of 5-repton long molecule.}
\label{netC}
\end{center}
\end{figure}

According to the Tab.\ref{classes}, for $35$ possible conformations of the 5-repton long molecule, $9$ classes allow for the correct description of the system. The probability of a particular class is equal to the sum of probabilities of states classified to this class. Here again, the obtained system can be described in terms of networks: nodes are identified with classes, and ties -- with connections between states belonging to a given class. Because of that, the obtained network is weighted. The network of classes is presented in Fig.\ref{netC}.

\subsection{Molecule in porous medium}
\begin{figure}[!hptb]
\begin{center}
\includegraphics[width=.4\textwidth, angle=0]{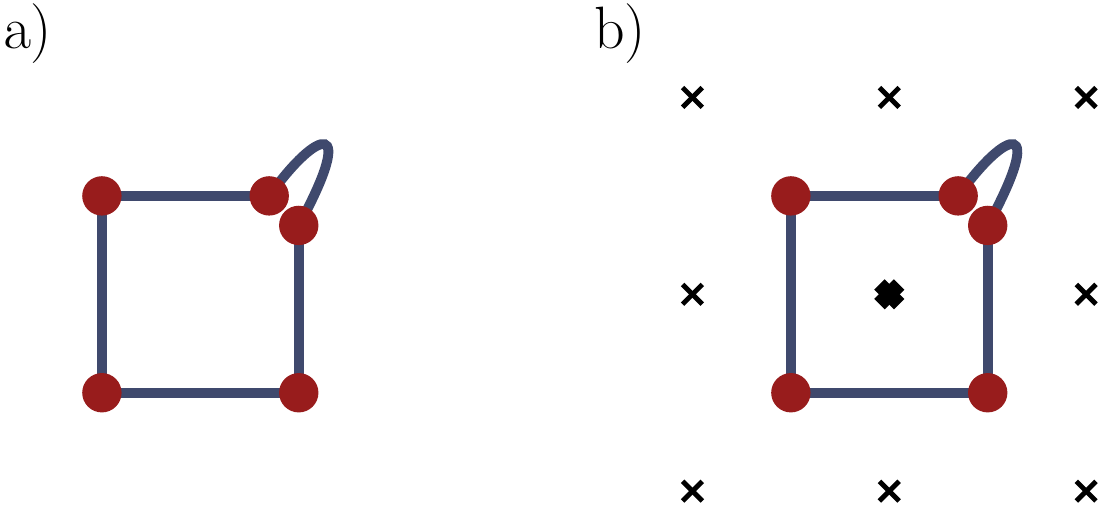}
\caption{a) One of the possible conformations of the 5-repton molecule. b) Some conformations becomes impossible if the molecules are embedded in a gel medium (the polymer is linked to a fibre with cross-section denoted by the solid bullet in the middle).}
\label{ogr}
\end{center}
\end{figure}

The first consequence of the presence of the medium is that not all theoretically possible molecule conformations are possible. An example is presented in Fig.\ref{ogr}, where one of the possible conformations of the 5-repton molecule is shown. The conformation presented in panel $a$ (and three similar conformations resulting from the space symmetry), is allowed. However, if we analyse molecules moving in the porous medium, such conformations cannot be obtained. The molecule encircles the fibre in the middle of  panel $b$ of Fig.\ref{ogr}. As we assume that both ends of each fibre are connected with the medium, this situation cannot happen. Because of that, some configurations have to be removed from the state space. The number of conformations which should be removed depends on the length of the molecule. In the case of five reptons, the four earlier mentioned conformations cannot occur. Because of that, the state space contains now $31$ states. What is also changed is the set of possible transitions between states, as the fibres of a gel act for migrating molecules as obstacles. From the originally specified set, we have to remove the transitions which pass across a fibre. An obvious consequence is the change of the number of classes. The result is presented in Tab.\ref{classesB}.

\begin{table}
{\tiny{
\begin{tabular}{c|*{6}{l}|c|c}
\textsc{class}&\multicolumn{6}{c|}{\textsc{classes given class can be transformed to}}&\textsc{degree}&\textsc{\#states}\\\hline
A1&C1&D1&&&&&2&12\\
B1&B1&D1&D2&&&&3&4\\
C1&A1&A1&D2&D2&&&4&6\\
C2&D2&D2&D2&D2&&&4&1\\
D1&A1&A1&A1&B1&D1&D2&6&4\\
D2&B1&C1&C1&C1&C2&D1&6&4\\
\end{tabular}
}}
\caption{Classes identified for the reduced state space of a circular molecule of 5 reptons in gel medium.}
\label{classesB}
\end{table}

The number of classes which initially equal to $9$ is reduced to $6$. The minimal degree is now equal to $2$, and the maximal one equals $6$. It results from the fact that now not only are some conformations not allowed but some repton displacements are also prohibited. Looking at Fig.\ref{przeS} we can see that transitions from the conformation in the middle, classified originally to the class $E1$, to the states classified to class $F1$ become impossible in the presence of gel fibres. As a consequence, the degrees of states and the number of classes decrease. One class from the original set of classes disappears as it contains states which are now impossible to obtain. Classes $A1$ and $B1$ of the original system are now merged, and the degree of states belonging to a new class is equal to $2$. Also original classes $E1$ and $F1$ are merged. Other classes remain unchanged with respect to states they include; what does change is the degree of states. The only exception is one-element class which consists of the state in which all reptons are placed in the same lattice cell. This class is identified in both cases. The number of possible transitions from this state in both cases is equal to $4$. 

\subsection{Molecule in porous medium and electric field}
The presence of an external electric field causes symmetry of the system to be partially lowered. As a consequence, the number of classes becomes higher than in the case without an electric field. In this case, the network of states is directed, with weights of transitions between given two states determined by the actual value and direction of the applied field. The class of states is now specified by their out-degree. Although, the number of links going in and out of each state is the same, their weights are different.\\
The number of classes depends on the direction of the applied field with respect to the square lattice. The results are presented in Tab.\ref{classesT}. If the direction of the field is parallel to the lattice axes, $21$ classes for the 5-repton long molecule are identified. A slightly smaller number of classes - $14$ - is identified for directions parallel to the diagonal of the lattice cells.% In this case some absorbing (trap) states are present and the minimal class out-degree is equal to $0$.
\begin{table}
{\tiny{
\begin{tabular}{c|c|c|c}
\textsc{\# links}&\textsc{\#classes $\vert\vert$}&\textsc{\#classes $\le$}&\textsc{\#states}\\\hline
2&7&4&12\\
3&3&2&4\\
4&5&4&7\\
6&6&4&8\\
\end{tabular}
}}
\caption{Classes identified for field directions parallel ($\vert\vert$) and diagonal ($\le$) to the lattice axis.}
\label{classesT}
\end{table}

\section{Probability distribution dynamics}
\label{dyn}
\subsection{Motion in field}
The possibility of the molecule's motion depends on the electric to thermal energy ratio and the direction of the applied electric field. Those two agents determine the probabilities of all possible shifts of the molecule in a given conformation. In the case of motion in the strong electric field, shifts which cause an increase of energy become practically impossible. The probabilities of those transitions are very close to zero. In this case, it is possible that for a given field direction some conformations may behave as traps. This situation occurs for the states for which all possible transitions are connected with a displacement of the repton in the direction opposite to the direction of the applied external electric field. A change of the molecule's position in such a conformation requires a very long time. We observe also that some states which are not traps themselves, quickly achieve a trap conformation as a result of the allowed transitions. However, for a majority 
of states a series of transitions leading to the molecule displacement in the space are possible. For moderate and weak fields, trap states are not observed, in the sense that probabilities of all potentially possible transitions are non-zero.\\
In all cases, the probabilities of transitions are calculated in accordance with the formula:
\begin{equation}
w(\varepsilon)=\dfrac{1}{1+exp(-\varepsilon\cos\alpha)}
\label{prob}
\end{equation}
where: $\varepsilon=\Delta E/kT$ - electric to thermal energy ratio, $\alpha$ - angle between the currently considered electric field vector and the direction in which a given repton changes its position.\\

\subsection{Stationary probability distribution of the molecule conformations}
In the case of the absence of any external force, the probabilities of all states are the same. Switching an external electric field on causes the differences in probabilities which depend on the field strength. For each case a transition matrix can be constructed. However, the problem may occur in the case of strong fields. The existence of absorbing (trap) states causes that corresponding columns in the transition matrix contain zeros. In this case, the probabilities of the states depend on the initial state. Taking this into account, we use the set of Master equations \cite{kamp}: 
\begin{equation}
\dfrac{dM_i(t)}{dt}=\sum\limits_{j\in S_i}M_j(t)w_{j\rightarrow i}-\sum\limits_{j\in S_i}M_i(t)w_{i\rightarrow j}
\label{de}
\end{equation}
where for a given state $i$ summing up goes through its neighbours $S_i$ with probabilities of transitions $w$ calculated for a given value and sense of the external electric field (in accordance with Eq.\ref{prob}). The stationary distribution is found as the asymptotic one for long times, with the probabilities of all allowed states initially equal.

\subsection{Exact enumeration and its continuous version for weighted networks}
Trap conformations may not be observed for all molecule sizes and not for all field directions. In the case of a 5-repton molecule, trap conformations are observed if the external field is directed along the diagonal of the lattice cells. For each of these four possible field directions, two trap conformations are identified. In this case one can evaluate the time after which, starting from any non-trap conformation, the molecule will be in the trap state. To find this time we use an approach similar to the exact enumeration method (EE) \cite{exen}.\\
The advantage of this method is that it allows for taking into account the whole set of initial conditions at the same time. In EE, a record is constructed of length equal to the number of possible states of the system. In the initial state, values for all non-trap states are set to one, and for the trap states to zero. Each element of the record keeps the information of the probability of a given state. At each step of EE procedure for all non-trap elements, a new value is calculated, while elements for trap states are for the whole time kept unchanged (with value equal to 0). The value currently assigned to each state is equally divided into all its neighbours. So a new value of each state is calculated as a sum of the fractions of probabilities obtained from states which can be transformed to this state. An important assumption in \cite{exen} is that all transitions are equally probable.\\
In our case this assumption cannot be maintained. Probabilities of transitions to different states are of  different values and they change with the external field direction and strength. To take this into account, here we propose a slightly different approach. Similarly to the original method, we construct a record of the size equal to the number of states. The initial values are set to $1$ for non-trap elements, and to $0$ for trap elements, with the latter values kept unchanged for the whole time. The difference relates to the method of calculating the new value for the non-trap states, which allows taking into account different probabilities of particular transitions. The idea is to use Master equations (Eq.\ref{de}). Such an approach reflects the different time scales of different transitions. The method proposed is a continuous version of the exact enumeration procedure for weighted networks. Below it will be used to evaluate the relaxation time to and from the absorbing states.

\section{Results}

Both the velocity and the diffusion coefficient are found from the stationary probability distribution of classes. There, the probability of each class is equal to the number of states represented by the class multiplied by the probability of each of these states. On the contrary, the results on the relaxation time are approximated. Below they are presented for the original and the reduced state space. 

\subsection{Velocity of the molecules in field}
For molecules migrating in the porous medium one can estimate the velocity of displacement as a function of $\varepsilon$. To obtain this quantity we calculate for each value of $\varepsilon$ the probability of each molecule conformation in the equilibrium state $P_i(\varepsilon)$. For this purpose we use the set of Eqs.\ref{de} and we find numerically the solution in asymptotically long time, with an initial state where each conformation is equally probable. Then we calculate the field dependence of the velocity, using the formula:
\begin{equation}
v(\varepsilon)=\sum\limits_{i=1}^NP_i(\varepsilon)\times \langle w_{i\rightarrow j}(\varepsilon)\times k\rangle_{j\in S_i}
\label{vau}
\end{equation}
where: $P_i(\varepsilon)$ - probability of a given state at a given value of $\varepsilon$, $w_{i\rightarrow j}(\varepsilon)$ - probability of a transition from state $i$ to state $j$ at a given field $\varepsilon$, $k\in(-1,0,1)$ - direction of a repton displacement in relation to a given external field direction.

\begin{figure}[!hptb]
\begin{center}
\includegraphics[width=.5\textwidth, angle=0]{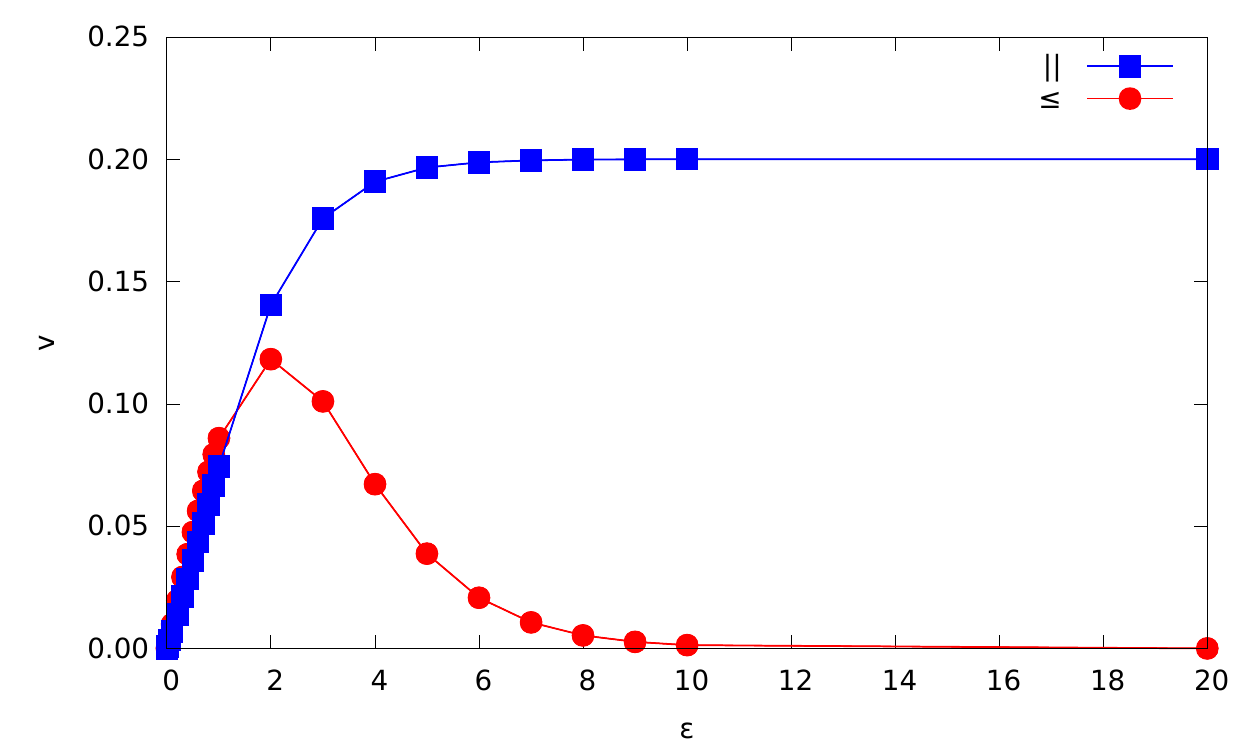}
\caption{Velocity as a function of $\varepsilon$ for the 5-repton long molecule. The line with squares presents results for the displacement direction parallel to the lattice axes. The line with dots presents results for the displacement direction at an angle $45^{\circ}$ to the lattice axes. Here $\varepsilon$ is unitless, temperature is set to 1, and the velocity is expressed in the number of cells per computer time step.}
\label{vOdE}
\end{center}
\end{figure}

The obtained result depends on the considered direction of the external electric field. The reason is that different behaviour is expected depending on the existence or lack of the absorbing trap conformations. Their presence will cause a decrease of the molecule velocity in the case of a strong electric field, as their probability in the equilibrium state then becomes dominant. Results for the 5-repton long molecule are presented in Fig.\ref{vOdE}.

The line with squares presents results obtained for the case free from the trap states, i.e. when the field is parallel to the lattice axes. In this case the velocity grows linearly with the growth of $\varepsilon$ for values of $\varepsilon\lessapprox 2$. Then the curve begins to saturate and reaches constant value for $\varepsilon\gtrapprox 5$.\\
If the molecules can get trap conformations, the results are qualitatively different. For $\varepsilon\lessapprox 1$ the curve (line with dots in Fig.\ref{vOdE}) behaves approximately in the same way, as for the field direction parallel to the lattice axes. In this range of $\varepsilon$, an increase of the velocity with field is observed. The influence of the trap states becomes visible for $\varepsilon\gtrapprox 2$, when a maximum of the velocity is found. For $\varepsilon\in(1,10)$ the velocity decreases, as $\exp(-\alpha\varepsilon)$. Accurate calculation of the $\alpha$ parameter is difficult due to the shape of the curve but it can be estimated at $\approx 0.5$. In \cite{buuu}, the same character of the field dependence of the velocity was found for linear molecules, with the coefficient $\alpha=3/2$.  For $\varepsilon\gtrapprox 10$ the calculated velocity achieves zero with numerical accuracy. 

\subsection{Diffusion coefficient in the presence of absorbing states}
Apart from the velocity, the diffusion coefficient as a function of $\varepsilon$ can also be estimated from the probabilities $P_i(\varepsilon)$ in an equilibrium state. We use the formula:
\begin{equation}
D(\varepsilon)\propto \dfrac{ \sum\limits_{i\in trap}P_i(\varepsilon)\sum\limits_{i\in not\;trap}P_i(\varepsilon)}{2}
\end{equation}

In other words, the diffusion coefficient is calculated as a product of the probability that the system is in the trap state (the first factor) and the probability that the molecule migrates in a media (the second factor). This is an approximation of the binary distribution with two states, where molecules move with constant velocity in one state and they are trapped in another state. 

\begin{figure}[!hptb]
\begin{center}
\includegraphics[width=.5\textwidth, angle=0]{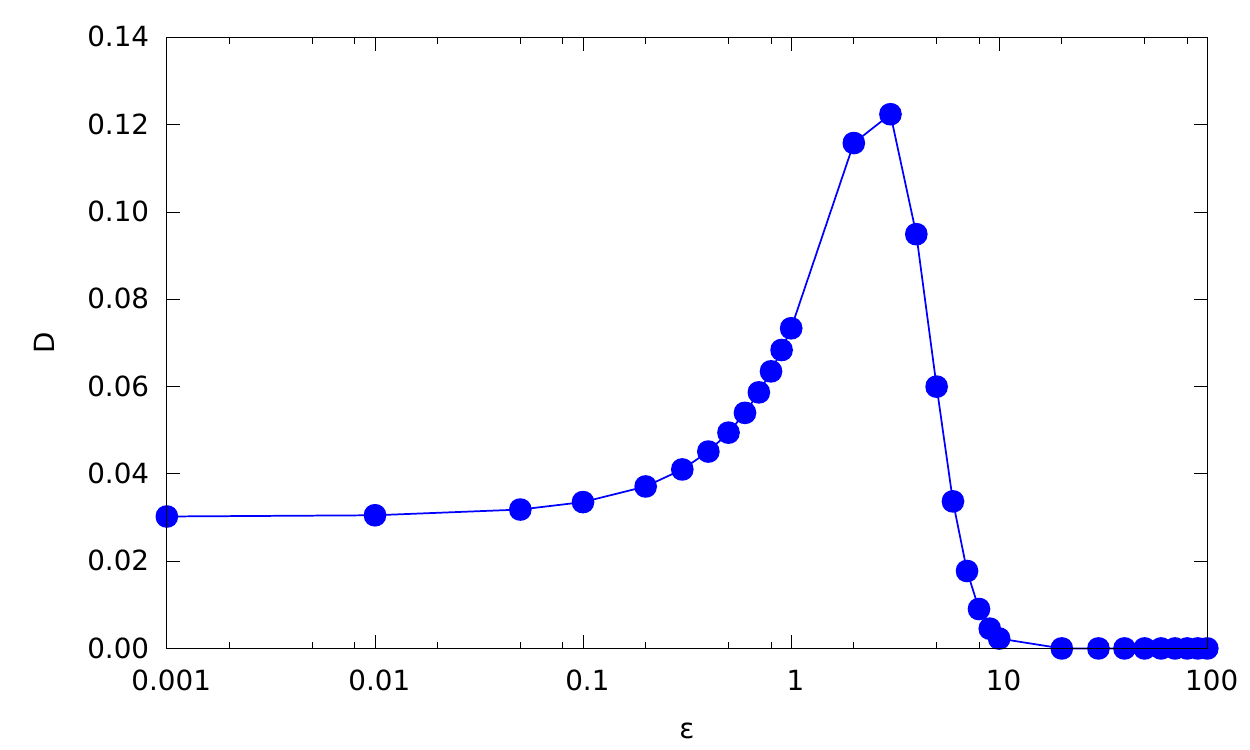}
\caption{Diffusion coefficient $D$ as a function of $\varepsilon$ for the 5-repton long molecule in the presence of absorbing states. Here $\varepsilon$ is unitless, temperature is set to 1, and $D$ is in lattice units squared per computer time step.}
\label{wd}
\end{center}
\end{figure}

The obtained results are presented in Fig.\ref{wd} in semi-logarithmic scale. The diffusion coefficient $D$ is very small and almost constant for $\varepsilon<0.1$, then increases linearly for $\varepsilon\in[0.1,2]$ and achieves maximum for $\varepsilon\approx 3$. In the range $\varepsilon\in(3,10)$ the diffusion coefficient decreases to $0$. Similarly to the velocity dependence, the value of the decrease parameter can only be estimated and is $\approx 0.43$. The shape of the curve is consistent with the velocity relation in the presence of trap states.

\subsection{Relaxation time to and from the absorbing states}
\label{extFi}
To estimate the relaxation time of the system we calculate the average value of $M$ of the whole set of non-trap states at each step of evolution. The method uses Master equations and the exact enumeration technique, as described in the last part of Section 3.3. As the result, we obtain the curve which decreases as $exp(-t/\tau)$. Parameter $\tau$ estimates the time needed to achieve trap states, against $\varepsilon$. The same formalism can be used to find the time needed for an escape from the trap conformations, provided that this process is possible in a finite time. As it was discussed earlier, this is possible if a shift caused by the thermal energy can occur against the field; this is the case in moderate and weak electric fields. In the case of a strong field, the molecule which achieves a trap conformation will remain in this state practically forever.\\

\begin{figure}[!hptb]
\begin{center}
\includegraphics[width=.5\textwidth, angle=0]{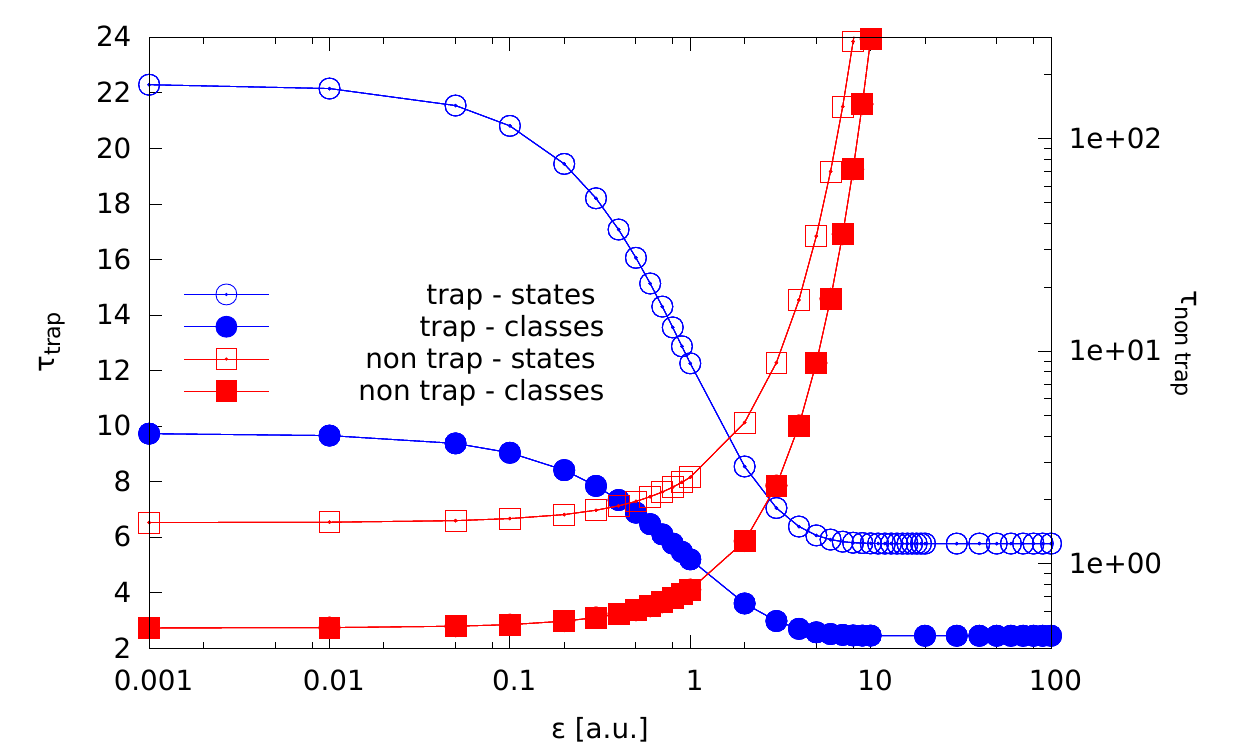}
\caption{Time of enrichment the trap state $\tau_{trap}$ (left hand side vertical axis) and time of escape from the trap $\tau_{non\;trap}$ (right hand side vertical axis) as a function of $\varepsilon$. The empty symbols refer to the states network, while the full one to the classes network. Here $\varepsilon$ is unitless, temperature is set to 1, and $\tau$ is in computer time steps.}
\label{czas}
\end{center}
\end{figure}

The obtained results for a 5-repton long molecule are presented in Fig.\ref{czas} in semi-logarithmic plot. The lines with dots show time $\tau_{trap}$ (left vertical axis) needed to achieve trap state (class of traps) as a function of $\varepsilon$. The empty symbols refer to the network of states, and the full ones to the network of classes. In a very low (below $\approx 0.1$) and in a high (above $\approx$ 5) range of $\varepsilon$, $\tau_{trap}$ does not change with the change of the field strength. In the first case the time of trapping is long, which is connected to possible thermal movements of reptons. In the case of a strong field it is clear that molecules will be quickly stopped in trap configurations. In between the two constant ranges the curve has a steep slope. An increase of the external electric field makes thermal shifts impossible, which shortens the time of trapping.\\
The line with squares shows time $\tau_{non\;trap}$ (right vertical axis) needed to escape from trap states (class of traps) as a function of $\varepsilon$. For values of $\varepsilon < 1$ the escape from the trap conformation occurs in a short time. When the field increases, the escape time begins to increase quickly and without an upper limit. For both curves, a qualitative agreement of results achieved for the network of states and the network of classes is obtained. Shorter relaxation times obtained in the case of the network of classes are the consequence of reducing the system size.

\section{Discussion}
We have demonstrated that the formalism of the reduction of the size of the state space of systems based on system symmetry can be applied in the case of state space of molecule conformations. The rate of the size reduction depends on the molecule length, and it is also different in the presence a of gel medium and the external electric field. In the last case the reduction of the system size is not as significant as in the absence of a field, because of the field-induced symmetry breaking. A qualitative agreement of the results of the relaxation time for the original network of states and for the reduced system of classes is obtained. It shows that the system properties are preserved in its compressed representation.\\

For the obtained states we show how probabilities of states can be calculated. The presented approach can be applied for any weighted system, even if absorbing states are observed in the system. We also show how the exact enumeration method can be generalized to be used in the case of weighted networks.\\
Although the results are presented for 5-repton long circular molecules, the proposed approach can also be applied for different lengths and shapes of molecules. Also in this case, some symmetries will be observed. The effects discussed above are expected to be present also for molecules of other shapes and lengths. In particular, in strong fields trap states are generic. Their absence in our case of a field parallel to the lattice axes seems to be an artifact of the lattice itself. The result that velocity decreases with the electric field strength seems to be paradoxical. However, a similar result was obtained in an experiment with DNA knots \cite{knot} and in another theoretical model for the linear polymers migrating in strong electric fields \cite{buuu}. Our results obtained for the reduced state space are consistent with literature.\\

We conclude that the method used to reduce the size of state space is general and useful numerically.\\

{\bf Acknowledgement:} 
The author is grateful to Krzysztof~Kulakowski for critical reading of the manuscript and helpful discussions. The research is partially supported within the FP7 project SOCIONICAL, No. 231288.

\end{document}